\documentclass{aa} 
\usepackage{graphicx}
\usepackage{times}
\usepackage{natbib} 
\usepackage{txfonts}
\bibpunct{(}{)}{;}{a}{}{,} 

\newcommand{\msun}{$\mathrm{{M}_\odot}$}

\newcommand{\mjup}{$\mathrm{{M}_J}$}
\newcommand{\mjups}{$\mathrm{{M}_J}$ }

\newcommand{\mearths}{$\mathrm{{M}_\oplus}$ }
\newcommand{\mum}{$\mathrm{\mu m}$}
\newcommand{\mums}{$\mathrm{\mu m}$ }

\newcommand{\eqref}[1]{(\ref{#1})}

\begin{document}

\title{Dust accretion onto high-mass planets} 
\author{Sijme-Jan Paardekooper\thanks{\emph{Present address:} Department of Applied Mathematics and Theoretical Physics, Wilberforce Road, Cambridge CB3 0WA, UK}} 
\institute{Leiden Observatory, Leiden University, Postbus 9513, NL-2300 RA Leiden, 
           The Netherlands \\
	   \email{S.Paardekooper@damtp.cam.ac.uk}} 

\date{Draft Version \today} 
  
\abstract{}{We study the accretion of dust particles of various sizes
    onto embedded massive gas giant planets, where we take into
    account the structure of the gas disk due to the presence of the
    planet. The accretion rate of solids is important for the
    structure of giant planets: it determines the growth rate of the
    solid core that may be present as well as their final enrichment
    in solids.} 
    {We use the RODEO hydrodynamics solver to solve the flow equations
    for the gas, together with a particle approach for the dust. The
    solver for the particles' equations of motion is implicit with
    respect to the drag force, which allows us to treat the whole dust
    size spectrum.}
    {We find that dust accretion is limited to the smallest particle
    sizes. The largest particles get trapped in outer mean-motion
    resonances with the planet, while particles of intermediate size
    are pushed away from the orbit of the planet by the density
    structure in the gas disk. Only particles smaller than
    approximately $s_\mathrm{max} =10$ \mums may accrete on a planet
    with the mass of Jupiter. For a ten times less massive planet
    $s_\mathrm{max}=100$ \mum. The strongly reduced accretion of dust
    makes it very hard to enrich a newly formed giant planet in solids.}{}
  
  
\maketitle

  
\section{Introduction}
Planets form in circumstellar disks consisting of gas and dust. Initially,
gas and solids are well-mixed with a dust-to-gas ratio of 1:100, which is 
similar to the interstellar value, and the dust particles will be as small as
interstellar grains ($\sim 1$ \mum). However, due to the pressure 
structure of the disk the dust particles start to move with respect to the
gas: the vertical component of the stellar gravity makes the particles
rain down onto the midplane of the disk, while the radial pressure gradient
of the gas pushes the particles inward \citep{1977MNRAS.180...57W}.

The magnitude of the velocity difference between gas and dust depends on the
particle size: small particles couple well to the gas and move slower than the
larger particles. These size-dependent velocities in the radial and vertical 
direction, together with Brownian motion and turbulence, will lead to 
collisions between the dust particles of various sizes. Assuming that impact 
velocities are low the particles stick together upon colliding, after which 
they continue as one large particle. This leads to a rapid depletion of small 
grains \citep{2005A&A...434..971D}.

However, sticking probabilities are still unclear, especially for larger 
particles. Experimental results \citep{1993Icar..106..151B,2000Icar..143..138B}
suggest that unless the relative velocities of particles are very low 
collisions lead to bouncing or even shattering of the particles. Another 
problem arises once the particles grow to sizes of approximately one meter,
because then the time scale for drag-induced radial migration becomes as short
as 100 yrs for particles at 1 AU \citep{1977MNRAS.180...57W}. These problems
led to a renewed interest in planetesimal formation through gravitational
instability \citep{1973ApJ...183.1051G,2002ApJ...580..494Y,
2004ApJ...603..292G}.
 
Once the planetesimals reach sizes of roughly 1 km the aforementioned problems
cease to exist. These bodies couple not well enough to the gas to experience
fast radial migration, and their gravitational influence is aiding their
growth. An important consequence of gravity-aided growth is that the heavier
bodies will grow fastest, which leads to a phase of oligarchic growth where
a few large bodies dominate the dynamics of the system 
\citep{1998Icar..131..171K,2000Icar..143...15K,2002ApJ...581..666K}. This
is an important epoch in planet formation, because it covers the phase from
planetesimals to planets of masses comparable to Mercury or Mars in the inner
regions of the solar system, while in the outer regions the full-grown cores
of giant planets may have been formed \citep{2006Icar..180..496C}. In the core
accretion model \citep{1996Icar..124...62P} of giant planet formation these 
cores of several \mearths attract significant amounts of gas from the nebula 
which results in the end in the birth of a gas giant planet.

The alternative for the core accretion model is the gravitational instability 
scenario, in which a massive disk becomes gravitational unstable and 
fragments into giant planets \citep{1997Sci...276.1836B,2004ApJ...609.1045M}.
However, whether a disk can become unstable depends sensitively on the
cooling time scale \citep{2000ApJ...529.1034P,2003ApJ...590.1060P} and it is
not yet clear if realistic protoplanetary disks will be subject to 
gravitational instabilities. Furthermore, the fragments produced in 
numerical simulations have masses around 10 Jupiter masses (\mjup), which
is rather high compared to the giant planets found in extrasolar planet
surveys. 

One of the key differences between the two scenarios is the presence of a
solid core. Unfortunately, the existence of solid cores in gas giant planets 
is still subject of debate. Models of the internal structure of Jupiter 
\citep{2004jpsm.book...35G} can not constrain the core mass from below due 
to uncertainties in the equation of state near the center of the planet. 
However, the same models do suggest that Jupiter is enriched in solids with
respect to the Sun. If Jupiter was formed by means of gravitational instability
it would have had to accrete these solids in a later stage.

Accretion of solids also determines the luminosity of forming planets, which
is important for future observations of protoplanetary disks. On
their way to the core the accreted planetesimals deposit their energy in
the gaseous envelope, which radiates away part of this energy into space. This
accretion rate plays a major role in the dynamics of forming giant planets
\citep{1996Icar..124...62P}.

It is therefore important to understand the process of accretion of solids
onto forming planets. \cite{1978Icar...35....1G} studied accretion onto solid 
cores, and the effect of gas drag was investigated by 
\cite{1985Icar...62...16W}, who found that inward moving planetesimals will
be captured in Mean Motion Resonances (MMRs). Only if the gas drag is strong
enough (i.e. the planetesimals are small enough) they will make it to the 
surface of the planet. At first sight one might think that this slows down
planetesimal accretion severely, but since their eccentricity is pumped up
in the resonances collisions become much more frequent and the small debris
resulting of catastrophic collisions subsequently accrete onto the planet
\citep{1985Icar...62...16W}. However, \cite{1993Icar..106..288K} showed that
even particles small enough to move through all resonances do not always reach 
the surface of the planet. Instead, they are transferred into inferior orbits
and they continue to move towards the central star.

These studies including gas drag assumed a uniform gas disk, but hydrodynamical
simulations of planets embedded in a gaseous disk show that planets more 
massive than $0.1$ \mjups start to restructure their environment 
\citep[e.g.][]{2002A&A...385..647D,2003ApJ...586..540D}. Eventually a deep
gap forms in the disk for the most massive planets $M_\mathrm{p} \ge$ \mjups 
\citep{1993prpl.conf..749L,1999ApJ...514..344B}. The pressure gradients 
associated with these gaps dramatically change the behavior of dust particles,
which may lead to a dust gap while there is no gas gap 
\citep{2004A&A...425L...9P}, and also to a substantially reduced accretion
rate of solids \citep{paard07}. However, the latter study was limited in
particle size and planet mass due to the fluid nature of the simulated dust
component. In this paper we use a particle-based method to study dust accretion
onto high-mass planets for arbitrary particle sizes. We focus on planets with
masses large enough to affect their direct environment, $M_\mathrm{p} \ge 0.1$ 
\mjups. This also allows us to do two-dimensional simulations, since the 
measured gas accretion rates for these planets in 2D are similar to the 3D 
values \citep{2003ApJ...586..540D}.

We start in Sect. \ref{dustaccsecEq} by reviewing the governing equations of 
motion, including the adopted gas drag law. In Sect. \ref{dustaccsecNum} we 
describe the numerical method used to integrate the equations of motion, and 
Sect. \ref{dustaccsecMod} is devoted to the disk model. In Sect. 
\ref{dustaccsecRes} we describe the results of the simulations, which we 
briefly discuss in Sect. \ref{dustaccsecDisc}. We conclude in Sect. 
\ref{dustaccsecCon}.  

  
\section{Equations of motion}
\label{dustaccsecEq}
Throughout we will work in a cylindrical coordinate frame $(r,\phi)$ with the 
central star in the origin. Because of the presence of a planet this is not an 
inertial frame, of which good use has been made in finding extrasolar planets, 
but we have to correct for this in the potential. The planet's orbit is 
circular, and the coordinate frame corotates with the planet at angular 
velocity $\Omega_\mathrm{p}$. Our unit of distance is the orbital radius of the
planet, which is then located at $(r,\phi)=(1,\pi)$ throughout the simulation.
The mass of the planet only enters the problem as a fraction of the stellar
mass $q=M_\mathrm{p}/M_*$. When quoting explicit planetary masses (relative to 
the mass of Jupiter, \mjup) we assume that the central star has a mass of 1 
\msun. Then 1 \mjups corresponds to $q=10^{-3}$.

\subsection{Gas}
The evolution of the gas component of the disk is governed by the Euler 
equations, which for this specific case are described in detail in 
\cite{paard06}. We do not solve the energy equation, but we use a locally
isothermal equation of state:
\begin{equation}
\label{dustacceqIso}
p=c_\mathrm{s}^2 \Sigma_\mathrm{g}
\end{equation}
where $\Sigma_\mathrm{g}$ is the surface density of the gas, $p$ is the
vertically integrated pressure, and the isothermal sound speed $c_\mathrm{s}$ 
is directly related to the disk thickness $H$:
\begin{equation}
\label{dustacceqSound}
c_\mathrm{s}=H~\Omega_\mathrm{K}=h~v_\mathrm{K}
\end{equation}
where $\Omega_\mathrm{K}$ is the Keplerian angular velocity, $h=H/r$ and
$v_\mathrm{K}=r\Omega_\mathrm{K}$. The gas has a kinematic viscosity $\nu$, 
which we take to be constant throughout the computational domain. We neglect
the self-gravity of the gas.

\subsection{Dust Particles}
The equations of motion for a dust particle read:
\begin{equation}
\frac{dv}{dt}=\frac{L^2}{r^3}-\frac{\partial \Phi}{\partial r} + f_r
\end{equation}
\begin{equation}
\frac{dL}{dt}=-\frac{\partial \Phi}{\partial \phi}+f_\phi
\end{equation}
where $v$ is denotes radial velocity, $L$ is the specific angular momentum, 
$\Phi$ is the gravitational potential and ${\vec f}$ is the drag force, which 
we specify below. The potential contains contributions from the central star,
the planet and indirect terms due to the acceleration of the coordinate
frame. We do not consider self-gravity for the dust particles.

\subsection{Drag Force}
The nature of the friction between gas and dust depends on the size of the 
dust particles relative to the mean free path of the gas molecules. This is
expressed by the Knudsen number:
\begin{equation}
Kn = \frac{\lambda}{2s}
\end{equation}
where $\lambda$ is the mean free path of the gas molecules and $s$ is the
size of the dust particles under consideration. When $Kn \gg 1$ we are in
the Epstein regime of free molecular flow, and the drag force is given by
\citep{schaaf}:  
\begin{eqnarray}
& &{\vec F}_{\mathrm{eps}} = -\pi s^2 \rho_\mathrm{g}\left|{\vec
    \Delta v}\right| 
{\vec \Delta v}\nonumber\\ 
\label{dustacceqSchaaf}
& & \left[\left(1+\frac{1}{m^2}-\frac{1}{4m^4}\right)
\mathrm{erf}(m) + \left(\frac{1}{m}+\frac{1}{2m^3}\right)
\frac{e^{-m^2}}{\sqrt{\pi}}\right]
\end{eqnarray}
where $\rho_\mathrm{g}$ is the gas density, ${\vec \Delta v}$ is the velocity 
of the dust particle relative to the gas, and $m$ is the relative Mach number 
of the flow: $m=|{\vec \Delta v}|/c_\mathrm{s}$. Equation 
\eqref{dustacceqSchaaf} has the following asymptotic behavior:
\begin{equation}
{\vec F}_{\mathrm{eps}} =\left\{ \begin{array}{ll}
       -\frac{\sqrt{128\pi}}{3}s^2 \rho_\mathrm{g} c_\mathrm{s} {\vec
	 \Delta v}~~~~  
       & \mbox{if~ $|{\vec \Delta v}| \ll c_\mathrm{s}$};\\
       -\pi s^2 \rho_\mathrm{g} |{\vec \Delta v}|{\vec \Delta v} 
       & \mbox{if~ $|{\vec \Delta v}| \gg c_\mathrm{s}$}. \end{array} \right.
\end{equation}
Because Eq. \eqref{dustacceqSchaaf} is difficult to implement numerically we 
interpolate between these limits in order to obtain the drag force for 
arbitrary velocities \citep{1975ApJ...198..583K}:
\begin{equation}
\label{dustacceqEps}
{\vec F}_{\mathrm{eps}} = -\frac{\sqrt{128\pi}}{3}s^2 \rho_\mathrm{g} c_\mathrm{s} f_\mathrm{D}
{\vec \Delta v}
\end{equation}
where 
\begin{equation}
f_\mathrm{D}=\sqrt{1+\frac{9\pi}{128}~m^2}
\end{equation}

When $Kn \ll 1$ the drag force is given by Stokes friction:
\begin{equation}
\label{dustacceqStokes}
{\vec F}_{\mathrm{sto}}=-6 \pi s k_\mathrm{D} \mu_{\mathrm{kin}}{\vec \Delta v}
\end{equation}
where $\mu_{\mathrm{kin}}$ is the kinematic viscosity of the gas, which we
write as: 
\begin{equation}
\label{dustacceqMukin}
\mu_{\mathrm{kin}}=\frac{1}{3}\rho_\mathrm{g} v_{\mathrm{th}}\lambda
\end{equation}
where $v_{\mathrm{th}}=\sqrt{8/\pi}c_\mathrm{s}$ is the mean thermal 
velocity of the gas. Note that $\mu_{\mathrm{kin}}$, and therefore also
${\vec F}_{\mathrm{sto}}$, is independent of the ambient gas density. This
in contrast with ${\vec F}_{\mathrm{eps}}$, for which a higher gas density
leads to a larger force of friction. 

In Eq. \eqref{dustacceqStokes} the drag coefficient $k_\mathrm{D}$ depends on the Reynolds number
$Re$ as follows:
\begin{equation}
\label{dustacceqKD}
k_\mathrm{D}=\left\{ \begin{array}{ll}
       1+0.15~Re^{0.687}~~~~ & \mbox{if $Re \leq 500$};\\
       3.96~10^{-6}~Re^{2.4} & \mbox{if $500 < Re \leq 1500$};\\
       0.11~Re & \mbox{if $Re > 1500$}. \end{array} \right.
\end{equation}
where 
\begin{equation}
Re=\frac{2 s \rho_\mathrm{g} |{\vec \Delta v}|}{\mu_{\mathrm{kin}}} =
3\sqrt{\frac{\pi}{8}} \frac{m}{Kn}
\end{equation}

For flows of intermediate Knudsen number reliable expressions for the drag
force are not readily available, so we just interpolate 
\citep[see][]{2003A&A...399..297W}:
\begin{equation}
{\vec F}=\left(\frac{3Kn}{3Kn+1}\right)^2{\vec F}_{\mathrm{eps}} + 
\left(\frac{1}{3Kn+1}\right)^2{\vec F}_{\mathrm{sto}}
\end{equation}
Using Eqs. \eqref{dustacceqEps}, \eqref{dustacceqStokes} and 
\eqref{dustacceqMukin} we can rewrite the total drag force as:
\begin{equation}
{\vec F}=-\sqrt{128\pi}~\frac{3Kn~f_\mathrm{D}+k_\mathrm{D}}
{\left(3Kn+1\right)^2}~s^2 \rho_\mathrm{g} c_\mathrm{s}Kn {\vec \Delta v}
\end{equation}
Upon dividing by the mass of the dust particle, $\frac{4}{3}\pi s^3 
\rho_\mathrm{p}$, where $\rho_\mathrm{p}$ is the internal density of a 
dust particle, we can write for the force per unit mass:
\begin{equation}
\label{dustacceqDrag}
{\vec f}=-\frac{\Omega_\mathrm{K}}{T_\mathrm{s}}{\vec \Delta v}
\end{equation}
where $\Omega_\mathrm{K}$ is the Keplerian angular velocity and $T_\mathrm{s}$ 
is the dimensionless stopping time:
\begin{equation}
\label{dustacceqStop}
T_\mathrm{s}=\sqrt{\frac{\pi}{8}}~
\frac{\left(3Kn+1\right)^2}{9Kn^2~f_\mathrm{D}+3Kn~k_\mathrm{D}}
~\frac{s}{r}~\frac{\rho_\mathrm{p}}{\rho_\mathrm{g}}~
\frac{v_\mathrm{K}}{c_\mathrm{s}}
\end{equation}
Note that in the limit $|{\vec \Delta v}| \ll c_\mathrm{s}$, $Re \ll 1$ Eq. 
\eqref{dustacceqDrag} is linear in the relative velocity of the particle, which has 
great advantages for numerically solving the equations of motion (see Sect. 
\ref{dustaccsecNum}). 

  
\section{Numerical method}
\label{dustaccsecNum}
\subsection{Gas}
We solve the flow equations for the gas using the RODEO method \citep{paard06}.
This method was extensively tested on the planet-disk problem, and also used
in two-fluid mode in \cite{2004A&A...425L...9P}. It makes use of an approximate
Riemann solver, together with stationary extrapolation 
\citep{1995A&AS..110..587E} to integrate the source terms. See \cite{paard06}
for details.

\subsection{Dust}
The dust component of astrophysical fluids can in principal be described
as being a continuous fluid or as a collection of particles. Both approaches
have their advantages and disadvantages. The major advantage for the fluid
approach is that the flow is accurately described even in regions of very
low dust density. In the particle approach these regions would be severely 
depleted of particles, and therefore the local resolution would be relatively
poor. In the context of embedded planets, the opening of a dust gap in
a gas disk \citep{2004A&A...425L...9P} should preferably be modeled using
a dust fluid.

However, the fluid approach is not valid in all circumstances 
\citep{2004ApJ...603..292G}. First of all,
one fluid element (one grid cell) should contain enough dust particles to 
define a density, and to work with averaged velocities. Second, there should
be enough interparticle collisions to make such an average meaningful. In a
gas-dominated disk this means that the particles should couple reasonably
well to the gas. In view of both these limitations of the fluid approach the
larger particles of the dust size distribution can not accurately be modeled 
as a fluid. The critical size depends on the gas density and the shape of the 
particles, but for dust grains larger than approximately 10 cm the fluid 
approach breaks down at 5 AU in the Minimum Mass Solar Nebula. If locally the 
gas density is lowered severely, for example by a gap-opening planet, this 
maximum size goes down correspondingly. 

Therefore in cases of low gas density the particle approach seems to be the
better option, although one has to make sure that enough particles reside
inside the gap in order to obtain accurate results. In this study we focus
on high-mass planets that open up gas gaps in the disk, and therefore we
chose for the particle approach. 

In order to integrate the equations of motion numerically we used a standard
second order symplectic integrator (leapfrog), in which alternatingly 
positions and velocities are updated. During a position update, velocities
are assumed to be constant and vice versa:
\begin{eqnarray}
\frac{d{\vec r}}{dt}&=&{\vec v},~~~\mbox{${\vec v}$ constant} \\
\frac{d{\vec v}}{dt}&=&{\vec a}({\vec r},{\vec v}),~~~\mbox{${\vec r}$ constant}
\end{eqnarray}
where ${\vec a}$ denotes the acceleration of the particle.

The position update is trivial. For the radial velocity update we need
to solve
\begin{equation}
\label{dustacceqMot}
\frac{dv}{dt}=\frac{L^2}{r^3}-\frac{\partial \Phi}{\partial r} -
\frac{\Omega_\mathrm{K}}{T_\mathrm{s}}\Delta v
\end{equation}
Because we keep the gas velocity constant, the same differential equation
applies to $\Delta v$. In order to make the integration scheme suited in 
the regime $T_\mathrm{s} \ll 1$ we treat the drag term implicitly. Formally this can
only be done in the limit $f_\mathrm{D}, k_\mathrm{D} = 1$ (subsonic and laminar regime). 
However, because supersonic drift velocities as well as high Reynolds numbers
only occur when $T_\mathrm{s} \gg 1$ this is not a problem, because then the effects
of gas drag are small and there is no need to treat the drag force implicitly.

Upon differencing Eq. \eqref{dustacceqMot} we obtain:
\begin{equation}
\frac{\Delta v - \Delta v_0}{\delta t}=
\frac{L^2}{r^3}-\frac{\partial \Phi}{\partial r} -
\frac{\Omega_\mathrm{K}}{T_\mathrm{s}}\Delta v
\end{equation}
which has the solution
\begin{equation}
\Delta v = \frac{\Delta v_0 + \delta t \left(
\frac{L^2}{r^3}-\frac{\partial \Phi}{\partial r} \right)}
{1+\frac{\Omega_\mathrm{K}\delta t}{T_\mathrm{s}}}
\end{equation}
A similar equation can be derived for the angular momentum equation. 
These equations are used to update the radial and angular velocities of the 
dust particles.

The method was tested in two limiting cases: $T_\mathrm{s}=\infty$ and $T_\mathrm{s} \ll 1$.
When there is no coupling to the gas ($T_\mathrm{s}=\infty$) the problem reduces
to the restricted three-body problem, which has the Jacobi constant as an
integral of motion:
\begin{equation}
\label{dustacceqJac}
J=E-{\vec \Omega}_\mathrm{p}\cdot {\vec L}
\end{equation}
where $E$ is the total energy of the particle under consideration. We reduced
the magnitude of the time step for the particle integration with respect
to the hydrodynamical time step until $J$ was conserved to 1 part in $10^6$
after $100$ orbits of the planet. This required 10 particle time steps per hydrodynamical time step. We have adopted this time step for all simulations presented in this paper.

\begin{figure}
\centering
\resizebox{\hsize}{!}{\includegraphics[bb=245 10 495 245]{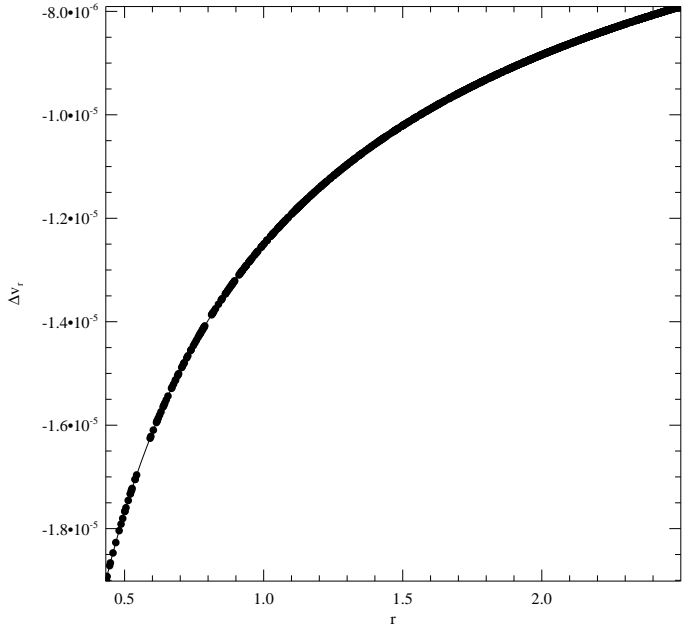}}
\caption{Radial velocity difference between gas and dust particles with
$T_\mathrm{s}=0.005$.}
\label{dustaccfig1}
\end{figure}

\begin{figure*}
\includegraphics[]{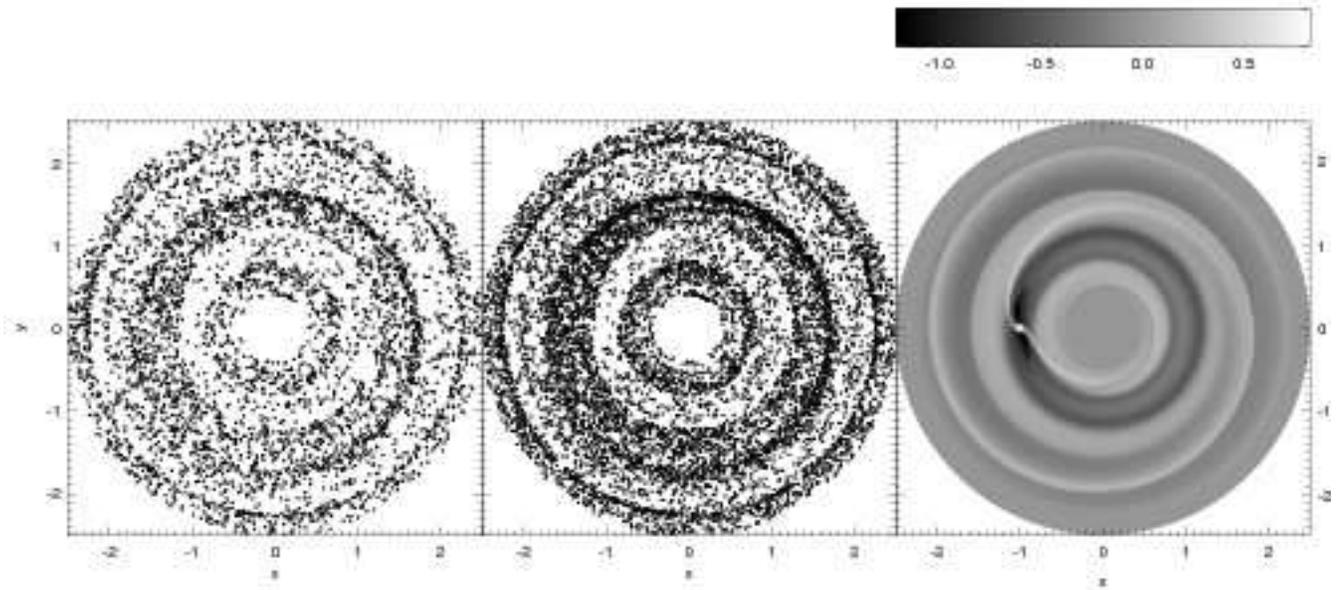}
\caption{Density structure after 50 orbits of a 1 \mjups planet around
  a solar mass star. 
Left panel: particle distribution ($T_\mathrm{s}=0$), $N=5000$. Middle panel:
particle distribution ($T_\mathrm{s}=0$), $N=10000$. Right panel: logarithm of the gas 
surface density.}
\label{dustaccfig2}
\end{figure*}

\begin{figure*}
\includegraphics[bb=42 10 525 242,width=\textwidth]{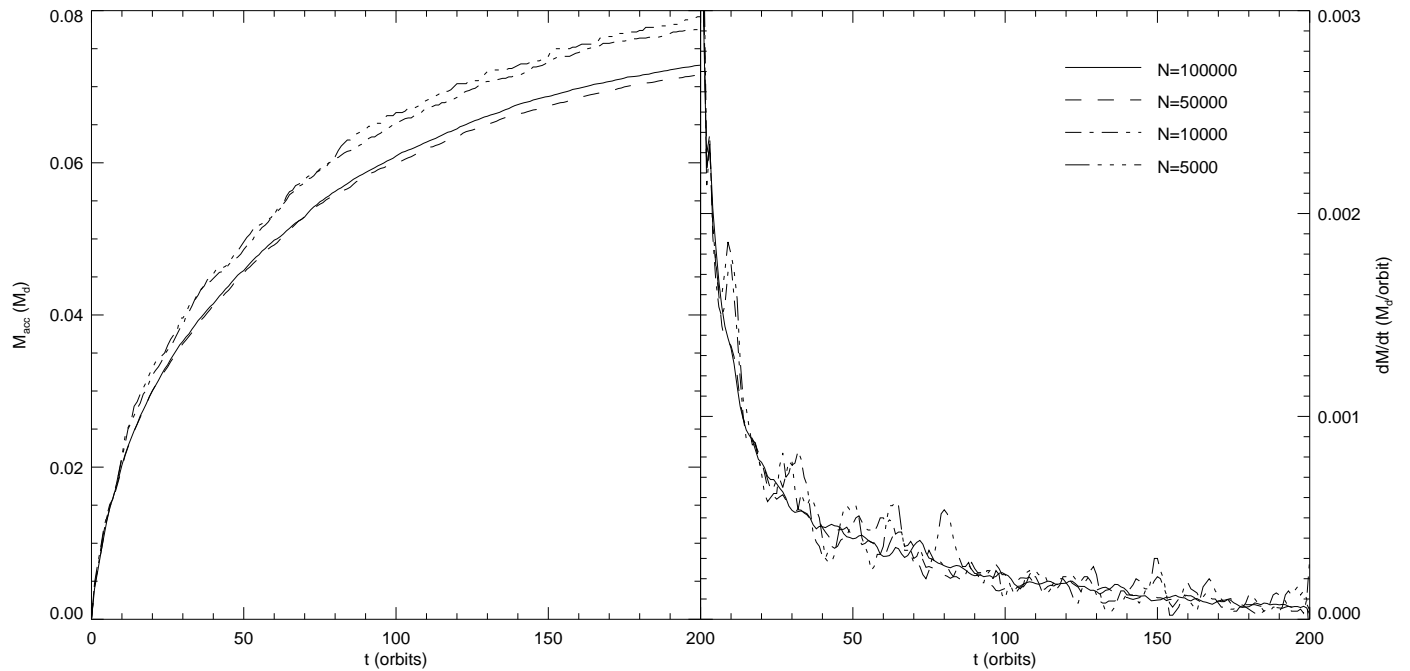}
\caption{Accretion of perfectly coupled particles onto a 1 \mjups planet
for four different amounts of particles. Left panel: accreted dust mass in 
units of the total dust mass in the disk.Right panel: corresponding 
accretion rates, in disk masses per orbit.}
\label{dustaccfig3}
\end{figure*}

In an axisymmetric disk without a planet, the radial drift velocity
is given by \citep{2002ApJ...581.1344T}:
\begin{equation}
\label{dustacceqDrift}
\Delta v = -\eta \left(T_\mathrm{s}^{-1}+T_\mathrm{s}\right) v_\mathrm{K}
\end{equation}
where $\eta$ is the ratio of the gas pressure gradient to the stellar gravity.
For uniform gas surface density, together with the temperature profile
dictated by Eq. \eqref{dustacceqSound}, $\eta=h^2$. We compare this analytical result
with the numerical solution for $T_\mathrm{s}=0.005$ in Fig. \ref{dustaccfig1}, and it turns 
out that they are indistinguishable. This shows that our implicit integration 
scheme for the drag force works correctly.

We model dust accretion onto the planet by taking away all particles within
one half of the Roche lobe $R_\mathrm{R}$ of the planet:
\begin{equation}
R_\mathrm{R}= \left(\frac{q}{3}\right)^{\frac{1}{3}}
\end{equation}
where $q$ is the ratio of the masses of the planet and the central star. 
The actual solid core is much smaller than this, so we do not measure the 
number of impacts onto this core directly. However, the high-mass planets that 
we consider in this paper are able to capture a dense gaseous envelope that 
is able to efficiently direct particles towards the center of the planet 
\citep[see also][]{paard07}. We do not take away gas from the computational 
domain, as has been done before to mimic gas accretion 
\citep{1999ApJ...526.1001L, 2002A&A...385..647D,paard06}. The process of 
accretion only influences the density structure close to the planet, and 
therefore this does not influence the results on dust accretion.

  
\section{Initial and Boundary Conditions}
\label{dustaccsecMod}
The gas disk model is the same as in \cite{2004A&A...425L...9P}. The
surface density of gas and dust particles is uniform initially, and we take 
the gas disk thickness to be $H=0.05~r$, so $h=0.05$, which is the canonical
value for simulations of planet-disk interaction. In order to compensate
for the radial pressure gradient dictated by Eq. \eqref{dustacceqIso} the gas orbits
at a slightly sub-Keplerian angular velocity, while the dust particles are 
on exact Keplerian orbits initially. We take all initial radial velocities 
of gas and dust to be zero. The gas has an anomalous viscosity $\nu$ that
is parametrized by the usual $\alpha$-prescription \citep{1973A&A....24..337S}:
\begin{equation}
\nu=\alpha~c_\mathrm{s}~H
\end{equation}
We adopt a constant value of $\nu=10^{-5}$, which corresponds to $\alpha=0.004$ 
at the location of the planet.

In order to calculate the stopping time in Eq. \eqref{dustacceqStop} we need a value
for the gas density. We fix the initial density to be $10^{-11}$ 
$\mathrm{g~cm^{-3}}$ at $r=1$, which is appropriate for the location of Jupiter 
in the Minimum Mass Solar Nebula. From the gas density we can calculate
the mean free path of gas molecules:
\begin{equation}
\lambda=\frac{m_{\mathrm{H_2}}}{\pi \rho_\mathrm{g} {r_\mathrm{H_2}}^2}
\end{equation}
where $m_{\mathrm{H_2}}$ and $r_\mathrm{H_2}$ are the mass and radius of an 
$H_2$ molecule, respectively. The internal particle density $\rho_\mathrm{p}$ 
is $1.25$ $\mathrm{g~cm^{-3}}$. Using these parameters, we can calculate $T_\mathrm{s}$ 
as a function of particle size $s$ (but note that $T_\mathrm{s}$ also depends on 
$|{\vec \Delta v}|$ through the factors $f_\mathrm{D}$ and $k_\mathrm{D}$).   

The dust density does not enter the equations of motion as long as it
remains much smaller than the gas density. If not, the gas is affected by the
drag force from the dust particles. We assume that initially the dust-to-gas
ratio is equal to 1:100, and that we therefore can ignore the feedback
on the gas. We vary the number of dust particles $N$ typically from $5000$ 
to $10^6$.

The computational domain extends from $r=0.4$ to $r=2.5$ in the radial
direction, and from $\phi=0$ to $\phi=2\pi$ in azimuth. This domain is 
covered by a uniform grid consisting of 128 radial and 384 azimuthal cells
which is used to evolve the gas component of the disk.

We take the boundary conditions for the gas to be non-reflecting 
\citep{1996MNRAS.282.1107G,paard06}, in order to keep the waves generated by
the planet from reflecting back into the computational domain. When a dust 
particle moves off the computational domain we remove it from the simulation. 

  
\section{Results}
\label{dustaccsecRes}
We have performed simulations with different particle sizes in order to
sample the parameter space in stopping time ranging from the limit of perfect
coupling $T_\mathrm{s}=0$ to the limit of no coupling $T_\mathrm{s}=\infty$. The former limit
offers an opportunity to compare the measured accretion rates to the gas 
accretion, so we will start by discussing perfectly coupled particles.

When we increase the size of the particles the stopping time increases 
as well. We can distinguish three regimes:
\begin{itemize}
\item{$T_\mathrm{s} \ll 1$: the motion of the particles is dominated by gas 
drag. However, significant gas-dust separation may occur in the presence of 
planets on time scales of approximately $\Omega_\mathrm{K}^{-1} 
T_\mathrm{s}^{-1}$ \citep[see][]{2004A&A...425L...9P}.}
\item{$T_\mathrm{s} \gg 1$: particle motion is dominated by gravitational 
interaction with the star and the planet, with the gas drag as a small 
perturbation.}
\item{$T_\mathrm{s} \approx 1$: the orbital time scale for a particle is 
comparable to the time scale for gas drag. Depending on the gas density, there 
may be region in which either gas drag or gravity will dominate.}
\end{itemize}
 
\subsection{Perfect coupling}
\label{dustaccsecPerfect}
First of all we discuss the case for $T_\mathrm{s} \ll \Omega_\mathrm{K} \Delta t$,
where $\Delta t$ is the magnitude of the time step as required for the 
integration of the Euler equations. For such a small value of the stopping
time the dust particles are forced to move with the gas velocity, and the
dust-to-gas ratio remains the same everywhere. In particular, gas and dust
accretion rates should be equal as well, when we correct for the factor
100 in the dust-to-gas ratio.

Gas accretion onto embedded planets has been studied before using different
numerical approaches \citep{1999ApJ...526.1001L,2002A&A...385..647D,paard06}, 
and they were found to agree reasonably well with each other. There is 
no strong dependence on the details of the accretion procedure 
(differences remain within $\sim$ 40 \%), except for intermediate mass planets
for which the Roche lobe is approximately equal to the disk scale height.
For these planets, accretion depends strongly on the conditions inside the 
Roche lobe and the measured accretion rates differ up to a factor of 2. 
Therefore we focus on a 1 \mjups planet, for which a well-defined gas
accretion rate has been measured of $\dot M = 10^{-4}$ disk masses per
orbit; a value that we would like to reproduce for the perfectly coupled
dust particles. Typically this value was reached within 200 orbits of the
planet.

We have varied the number of particles in the simulation from 5000 to
$10^5$. Note that in order to accurately measure an accretion rate the 
number of particles that is accreted onto the planet every orbit should be 
larger than one. For an accretion rate of $10^{-4}$ disk masses per orbit 
the total number of particles in the simulation $N$ therefore needs to be 
larger than $10^4$. For a smaller value of $N$ we need to rebin the accreted
particles into larger time intervals. The loss of time resolution is of no
importance for the final accretion rate because it does not vary rapidly
after $\sim$ 50 orbits.

In Fig. \ref{dustaccfig2} we show the particle distribution for a simulation of a 
1 \mjups planet after 50 orbits. In the gas density (right panel) we see
the gap that is developing, as well as the prominent spiral waves excited
by the planet. For a run with 10000 particles (middle panel) we can clearly
see the gap and also the spiral waves in regions where the particle density 
is high enough. However, in the gap region near the planet, where in the 
gas density we see the origin of the spiral waves, nothing is to be seen in 
the particle density. This is because the spatial resolution in a particle 
simulation critically depends on the particle density. In regions of low
density, as in the case of a disk gap, the resolution is correspondingly low 
and the spiral waves are not visible. This is of no concern for the outcome
of these simulations, however, because the dust density has no dynamical 
effect. If the density was used for example in calculating the number of
collisions between these particles, which in turn would create smaller dust
grains, resolution effects would play an important role. The same would be 
true if the dust density becomes high enough that the gas feels its drag. The 
only thing we need to worry about for these simulations is that we accrete 
enough particles per time step (see above).

Even if we lower the number of particles $N$ to 5000 there is still density 
structure visible (left panel of Fig. \ref{dustaccfig2}). The outer edge of the
gap is less sharp than in the case of $N=10000$, but all features in the 
density can still be identified.

Also, if we look at the accretion rates for different values of $N$ in Fig. 
\ref{dustaccfig3} the results look promising, as expected. After $200$ orbits, the 
amount of disk mass 
that was accreted for the various number of simulated particles differs less 
than 15 \%. Even more, the final accretion rates agree very well, yielding
a mean value of $\dot M=9.5~10^{-5}$ disk masses per orbit, in close agreement
with the value that came out of gasdynamical simulations. The total mass that
is accreted as well as the slightly lower accretion rates can be attributed
to the large imposed accretion inside the Roche lobe. While for the gas case
one usually takes out a small \emph{fraction} of the density away from the
accretion region every time step, for our particle simulations we remove
\emph{all} particles from the inner half of the Roche lobe. Such a high 
imposed accretion rate tends to decrease the total amount of mass that is 
accreted \citep[see Fig. 7 of][]{paard06}.

We conclude that we are able to reproduce the gas accretion rates in the limit
of perfectly coupled particles, and that we only need a relatively small amount
of particles to achieve this.  

\begin{figure}
\centering
\resizebox{\hsize}{!}{\includegraphics[]{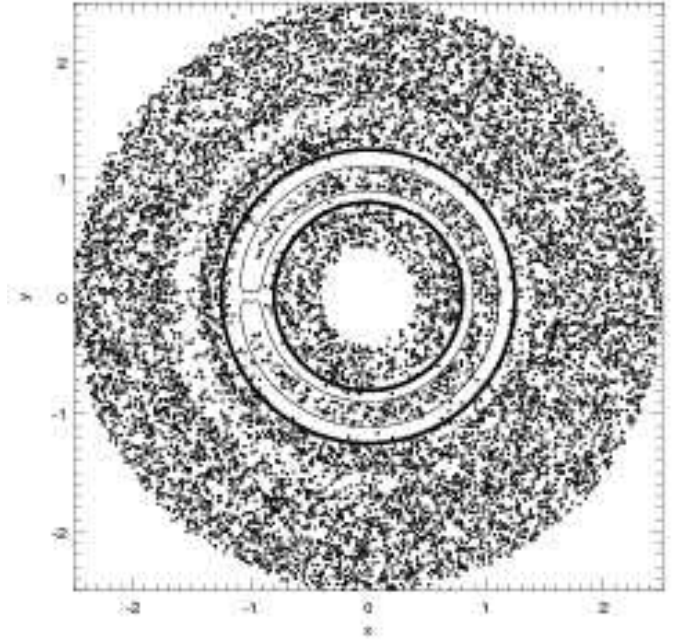}}
\caption{Particle distribution ($N=10000$) for $T_\mathrm{s}=\infty$
  after 500 orbits of a 1 \mjups planet. Thick contours mark the
  boundary of the accretion region of the planet. The thin contour
  gives the corresponding zero-velocity curve.} 
\label{dustaccfig4}
\end{figure}

\subsection{No coupling}
\label{dustaccsecNocoup}
In this section we will focus on the limit of no coupling to the gas. 
For these particles the problem is equivalent to the restricted three-body
problem, and therefore we can gain some insight by examining the integral of
motion of the problem, $J$ (see Eq. \eqref{dustacceqJac}). In particular, the 
motion of a particle is confined to a region bound by so-called zero-velocity 
curves, which can be obtained by solving Eq. \eqref{dustacceqJac} for a given 
value of $J=J_0$ and all velocities equal to zero. Particles with a starting 
value $J=J_0$ will never cross the zero-velocity curve associated with $J_0$. 
This means that the region in the disk from which the planet may accrete
particles is also bounded, and once this feeding zone is empty accretion 
will stop. 

In Fig. \ref{dustaccfig4} we show the particle distribution after 500 orbits of
the planet for $T_\mathrm{s}=\infty$ and $N=10000$. The region close to the 
planet is almost empty, and two annular empty rings appear at approximately 
$r=1.1$ and $r=0.9$. Particles in these rings travel on horseshoe orbits and 
suffer close encounters with the planet, which leads to accretion of these 
particles. Material that orbits further away from the planet ($r>1.1$ and 
$r<0.9$) can not reach the planet because of the constraint set by the Jacobi 
integral. The thick contours mark the radii at which particles may \emph{just} 
make it to the planet, which is shown by the zero-velocity curve (thin contour)
corresponding to the Jacobi constant at the location of the thick contour. 

The thick contours trace the outer edges of the feeding zone reasonably well,
and almost all particles initially located just inside this feeding zone
are accreted by the planet. However, particles closer to $r=1$ do not 
suffer close encounters with the planet: they move on stable tadpole orbits
and will not be accreted. In the solar system similar objects can be found
around the orbit of Jupiter: the Trojan asteroids. The end result of the 
simulation is a particle distribution with two empty ''rails'' at a radial 
distance of $0.1$ from the planet, together with a non-empty corotation region.
The empty semi-circle outside the orbit of the planet is due to resonant 
perturbations of the 2:1 outer mean motion resonance.

\begin{figure}
\centering
\resizebox{\hsize}{!}{\includegraphics[bb=250 10 495 240]{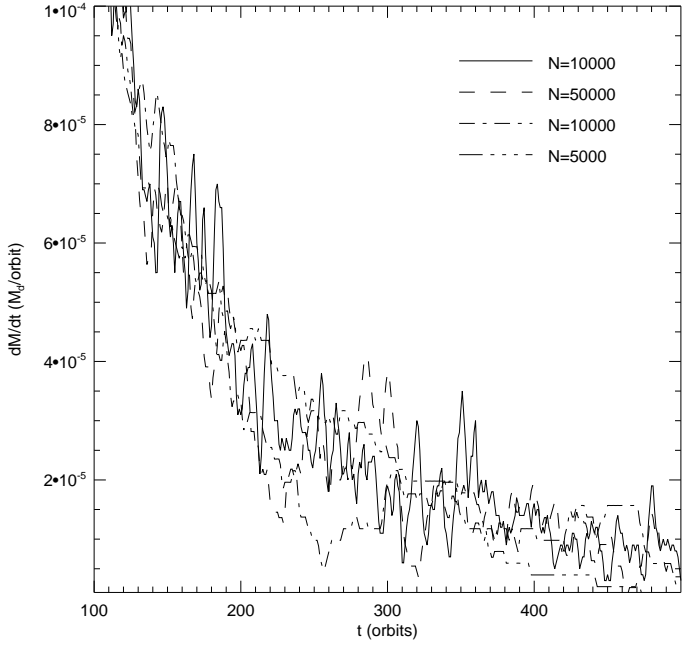}}
\caption{Accretion of uncoupled particles onto a 1 \mjups planet
for four different amounts of particles.}
\label{dustaccfig5}
\end{figure}

In Fig. \ref{dustaccfig5} we show the accretion rate up to 500 orbits for the 
same simulation, for four different values of $N$. Before $t=100$ (not shown in
the figure) the accretion rate is very high, and the planet accretes roughly
10 \% of the total disk during that time. However, after approximately
$120$ orbits the accretion rate drops below the gas accretion rate ($10^{-4}$
disk masses per orbit) and it continues to decline for the whole simulation 
time. After $500$ orbits the particle accretion rate is an order of magnitude
below the gas accretion rate, which makes further dust accretion negligible.

\begin{figure*}
\includegraphics[bb=42 10 525 242,width=\textwidth]{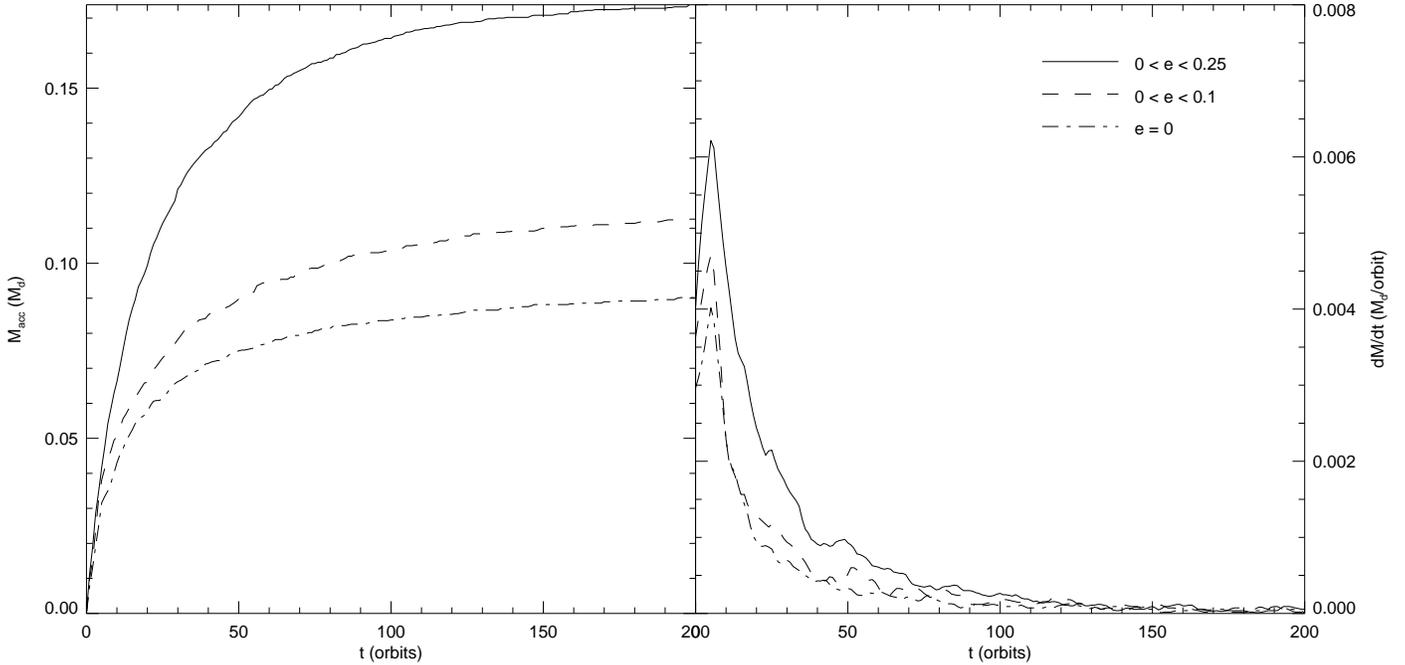}
\caption{Accretion of uncoupled particles ($N=5000$) onto a 1 \mjups planet
for three different initial eccentricity distributions. Left panel: accreted 
dust mass in units of the total dust mass in the disk. Right panel: 
corresponding accretion rates, in disk masses per orbit.}
\label{dustaccfig6}
\end{figure*}

All particles were started on circular orbits until now, while mutual 
gravitational interactions between larger boulders leads to eccentricity
excitation \citep{1990Icar...87...40G,1992Icar..100..440G}. The eccentricity
of the particles' orbits determines the width of the accretion zone, and
therefore the total mass that may be accreted. However, because the feeding
zone is still bounded the accretion rate will still tend to zero after a 
certain amount of time. This is illustrated in Fig. \ref{dustaccfig6}, where we
show the accretion rates for three different initial eccentricity 
distributions. All eccentricities in the specified range were uniformly 
distributed amongst the particles. It is clear that from approximately 
$130$ orbits the accretion rate is the same for all initial eccentricities,
which is the same time scale as for the clearing of the feeding zone (see
above). Therefore a non-zero initial eccentricity for the planetesimals
enlarges the accretion zone, and therefore increases the total amount
of mass that can be accreted (see the left panel of Fig. \ref{dustaccfig6}),
but it does not affect the time scale on which the feeding zone is cleared.

\subsection{Steady gas disk}
Clearly, the feeding zone needs to be replenished one way or the other in 
order for accretion to continue. In this section, we explore the possibility
that gas drag may prevent the feeding zone from getting empty. We do this 
by keeping the gas disk steady, in order to separate the effects on accretion
of a non-uniform gas disk from effects due to pure dust dynamics. Also this 
gives us an opportunity to compare with previous work by 
\cite{1993Icar..106..288K}. 

Based on Eq. \eqref{dustacceqDrift} one might expect that dust accretion would
always be stronger than gas accretion. Because the mass that flows inward 
through a circle with radius $r$ equals:
\begin{equation}
\label{dustacceqMdot}
\dot M = -2\pi~r \Sigma v_r
\end{equation}
the ratio of gas and dust accretion rates would be given by:
\begin{equation}
\label{eqMdotrel}
\frac{\dot M_\mathrm{d}}{\dot M_\mathrm{g}}=
\frac{2\pi~r\Sigma_\mathrm{d}v_{r,\mathrm{d}}}
{2\pi~r\Sigma_\mathrm{g} v_{r,\mathrm{g}}}=
d \left(1+\frac{\Delta v}{v_{r,\mathrm{g}}}\right)
\end{equation}
where $d$ is the dust-to-gas ratio and $\Delta v$ is given by Eq. 
\eqref{dustacceqDrift}. In an unperturbed disk with constant surface density and 
kinematic viscosity $\nu$ the radial velocity of the gas is equal to 
$v_{r,\mathrm{g}}=-\frac{3\nu}{2r}$, and therefore we have:
\begin{equation}
\frac{\Delta v}{v_{r,\mathrm{g}}}=
\frac{h^2 \left(T_\mathrm{s}^{-1}+T_\mathrm{s}\right) 
v_\mathrm{K}}{\frac{3}{2}\alpha c_\mathrm{s} h} =
\frac{2 \left(T_\mathrm{s}^{-1}+T_\mathrm{s}\right)}{3 \alpha}
\end{equation}
When all dust grains have the same stopping time $T_\mathrm{s}$ the gas-dust mixture
that is accreted by the planet has an effective dust-to-gas ratio of
\begin{equation}
d^*=d\left(1+\frac{2 \left(T_\mathrm{s}^{-1}+T_\mathrm{s}\right)}{3 \alpha}\right)
\end{equation}
Using $\alpha=0.004$, already for $T_\mathrm{s}=0.001$ the planet would get enriched
in solids by more than 15 \%.

However, the planet may not be able to grab all the material that is passing
by \citep{1993Icar..106..288K} which leads to lower dust accretion rates.
Gas, however, may be accreted at a \emph{higher} rate than predicted by
Eq. \eqref{dustacceqMdot} \citep[see][]{1999ApJ...526.1001L,2005astro.ph.12292L}.
It is therefore not clear, even in an unperturbed gas disk, how dust accretion
compares to gas accretion. 

\begin{figure}
\centering
\resizebox{\hsize}{!}{\includegraphics[bb=255 10 510 260]{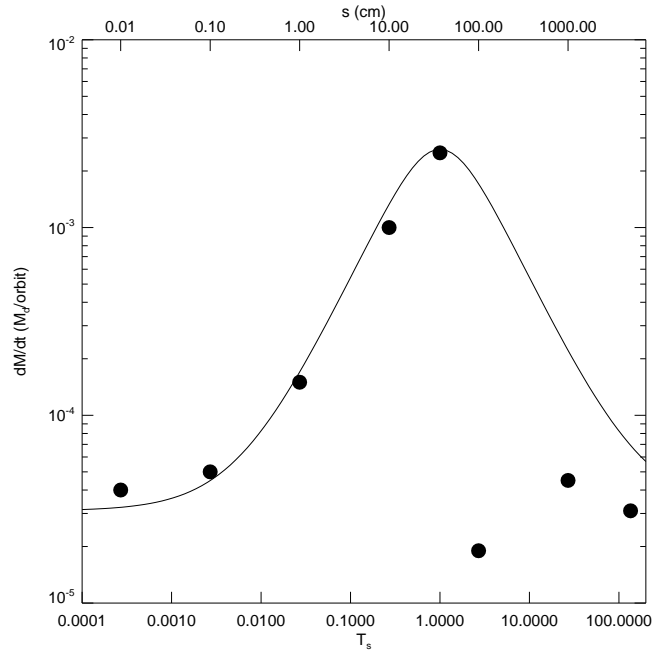}}
\caption{Accretion of particles onto a 1 \mjups planet. The solid line
gives the accretion rate of dust particles predicted by
Eq. \eqref{eqMdotrel}. All runs included 5000 particles.} 
\label{dustaccfig7}
\end{figure}

In Fig. \ref{dustaccfig7} we show the dust accretion rates of dust particles 
onto a 1 \mjups planet as a function of particle size. Note that we do not take
into account effects of the particle size distribution in the disk: the 
accretion rates shown would be valid if \emph{all} particles in the disk 
would have a single size. The solid line gives the accretion rate predicted 
by Eq. \eqref{dustacceqMdot} with the radial velocity of Eq. 
\eqref{dustacceqDrift}. Towards the lowest values of $s$ (and therefore the 
lowest values of $T_\mathrm{s}$) it approaches the value of the gas accretion. 
Note, however, that this is not the \emph{measured} gas accretion rate, which 
is higher due to the high accretion efficiency of the planet 
\citep{1999ApJ...526.1001L}.

Particles $T_\mathrm{s} \ll 1$ have a small radial velocity, and they can essentially 
all be captured by the planet. This results in Fig. \ref{dustaccfig7} in accretion
rates close to the predicted value for the smallest particles. When $T_\mathrm{s}$ 
increases towards 1, the radial velocity of the dust particles increases, but
at the same time is is easier for the planet to accrete them because they 
suffer no strong gas friction. Therefore essentially all particles with 
$T_\mathrm{s} < 1$ can be accreted by the planet.

\begin{figure}
\centering
\resizebox{\hsize}{!}{\includegraphics[bb=260 10 495 240]{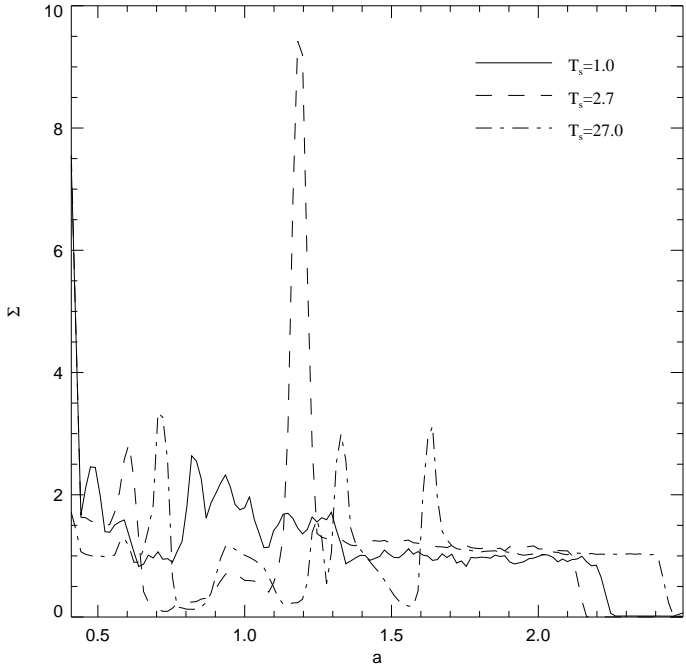}}
\caption{Particle surface density after 50 orbits of a 1 \mjups planet and 
three different values for $T_\mathrm{s}$. }
\label{dustaccfig8}
\end{figure}

\begin{figure}
\centering
\resizebox{\hsize}{!}{\includegraphics[]{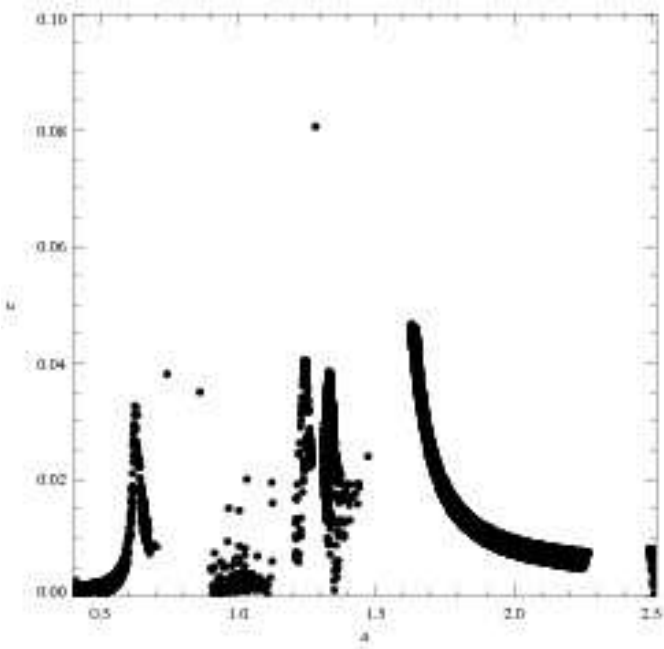}}
\caption{Particle semi-major axis versus eccentricity after 200 orbits of a 
1 \mjups planet and $s=1000$ cm ($T_\mathrm{s} \approx 27.0$).}
\label{dustaccfig9}
\end{figure}

For larger particles a dramatic change in accretion rate occurs. Increasing
the size of the particles by less than a factor of three leads to a reduction
of accretion by almost two orders of magnitude. The reason for this low 
accretion rate lies in the mechanism of resonance trapping 
\citep{1985Icar...62...16W}, where inward moving particles are captured in
mean motion resonances (MMRs). The drag force that moves the particles inward 
is exactly balanced by resonant perturbations which are directed away from 
the planet. This is illustrated in Fig. \ref{dustaccfig8}, where
we show azimuthally averaged particle surface density as a function of 
particle semi-major axis $a$. Because the eccentricities remain below $0.1$
(see Fig. \ref{dustaccfig9}), $a$ almost coincides with $r$. 

The solid line in Fig. \ref{dustaccfig8} corresponds to the particles size responsible
for the peak of the accretion rate in Fig. \ref{dustaccfig7}, $s=37.0$. There is
a lot of structure near the position of the planet, but no gap is forming and
there is no structure in the outer disk. Note that the surface density goes to 
zero at $a \approx 2.25$ because of the movement of the particles towards the 
central star. In the inner disk a small feature is visible at $a=0.48$, which 
we can attribute to the $1:3$ MMR. Note, however, that these structures in the
\emph{inner} disk are not stable: resonant perturbations as well as gas drag 
both push the particles inward. 

When we increase the particle size by less than a factor of three (dashed line
in Fig. \ref{dustaccfig8}) the situation is remarkably different. Near $a=1$ an 
annular gap starts to form, together with a strong overdensity at $a=1.2$, 
which is close to the 4:3 mean motion resonance ($a \approx 1.21$). Apart from
this large peak there is no structure in the outer disk. The inner 
disk shows a feature near the 1:2 MMR at $a=0.63$. 

Increasing the particle size by another factor of 10 again changes the picture
(dash-dotted line in Fig. \ref{dustaccfig8}). Now the feature at $a=1.2$ has 
disappeared and instead we can see a peak near the 2:1 MMR at $a=1.6$, 
together with a feature at $a=1.3$, which corresponds to the 3:2 MMR.
In the inner disk the 2:3 MMR is excited at $a=0.76$. 

It is clear from Fig. \ref{dustaccfig8} that resonance trapping occurs when $T_\mathrm{s} > 1$,
and that weaker-coupled particles may occupy lower-order resonances. Close to
a resonance the eccentricity of a particle is excited. In Fig. \ref{dustaccfig9} we 
show the eccentricity $e$ of the particles' orbits for
the $T_\mathrm{s}=27.0$ case from Fig. \ref{dustaccfig8}. At the location of the 2:1 MMR $e$
is excited to a maximum of approximately $0.05$, which is consistent with
the analytical result of \cite{1985Icar...62...16W} who showed that the
equilibrium eccentricity for a particle in a $(j+1):j$ MMR is approximately
given by $\bar{e}=0.07/\sqrt{j+1}$.

Resonance trapping occurs because resonant perturbations tend to change the 
semi-major axis of the particle's orbit in such a way that it moves away 
from the planet. For outer resonances, this means that this effect opposes the 
drag-induced inward migration. Inner resonances can never trap particles in
a stable way, unless the drag force is directed outward. This may happen when
a strong gas density gradient is present, possibly at the outer edge of a
gas gap in the disk. 

The strength of a resonance $(j+1):j$ depends on the mass planet-to-star mass 
ratio $q$ as well as on the order of the resonance $j$. Higher order 
resonances, which are located closer to the planet, are stronger and they
can stop inward moving particles more easily. Therefore the location at which
a particle is stopped by resonant perturbations depends on the particle size
$s$. This is clear from Fig. \ref{dustaccfig8}, where particles with $T_\mathrm{s}=27.0$ 
($s=1000$ cm) become trapped in the $j=1$ resonance, while particles with
$T_\mathrm{s}=2.7$ ($s=100$ cm) move all the way to the $j=3$ resonance. 
\cite{1985Icar...62...16W} derived a minimum size for a particle that is 
able to move through the $j^{\mathrm{th}}$ resonance:
\begin{equation}
\label{dustacceqSmin}
s_{\mathrm{min}}=\frac{\rho_\mathrm{g} h r_\mathrm{p}}
{3\rho_\mathrm{p}~q~C(j)~j^{3/2}}
\end{equation}
where $C(j)$ is an increasing function of $j$. For particles smaller than 
$s_\mathrm{min}$ the drag force is stronger than the resonant perturbations
and the particles are able to move inward through the resonance. Because
$s_\mathrm{min}$ becomes smaller for higher values of $j$ they may get trapped
in a resonance closer to the planet.

For higher values of $j$, successive resonances are more closely spaced and 
their overlap may lead to chaotic behavior \citep{1980AJ.....85.1122W}.
The distance at which this happens is approximately 
\citep{1989Icar...82..402D}:
\begin{equation}
\label{dustacceqAmax}
\left|r-r_\mathrm{p}\right| \approx 1.5~q^{2/7}
\end{equation}
For a planet of 1 \mjups this means that particles cannot be trapped closer
than $\left|r-r_\mathrm{p}\right| = 0.2$. It is interesting to note that this
is also approximately equal to the width of the gas gap that is opened by
a planet of the same mass \citep[see][]{1999ApJ...514..344B,paard06}.

\begin{figure}
\centering
\resizebox{\hsize}{!}{\includegraphics[bb=250 10 530 240]{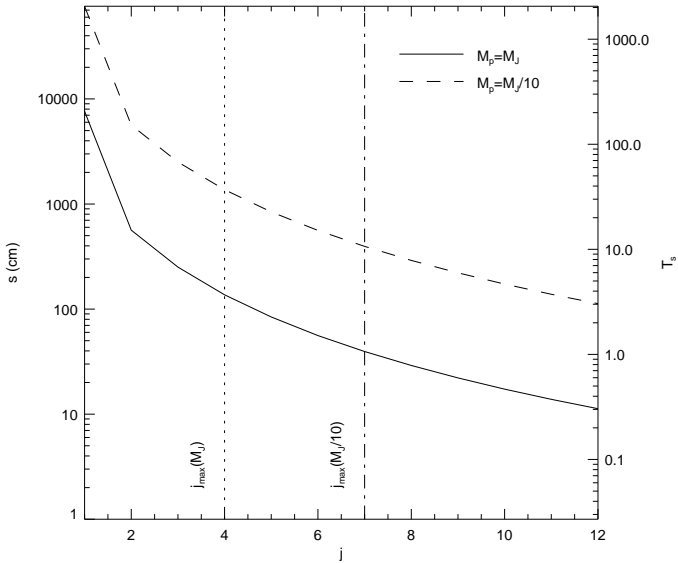}}
\caption{Minimum particle size that can be trapped in the $j^{\mathrm{th}}$
resonance, for a 1 \mjups planet and a $0.1$ \mjups planet. The vertical lines
mark the onset of chaotic orbits for high $j$.}
\label{dustaccfig10}
\end{figure}

In Fig. \ref{dustaccfig10} we show $s_\mathrm{min}$ as a function of $j$ for a
1 \mjups planet and a $0.1$ \mjups planet. Also shown are the maximum values 
of $j$ for which no chaotic orbits exist (from Eq. \eqref{dustacceqAmax}). 
Focusing on the 1 \mjups planet we see that only particles smaller than 
approximately $100$ cm are able to travel through the $j=4$ resonance, beyond 
which no stable trapping exists because of Eq. \eqref{dustacceqAmax}. In terms 
of accretion onto the planet this means that only particles smaller than $100$ 
cm are sufficiently coupled to the gas to make it all the way to the surface of
the planet. From Fig. \ref{dustaccfig7} we see that indeed the large jump in 
accretion rate happens around $s=100$ cm. Our numerical result for 
$s_\mathrm{min}$ agrees within a factor of 2 with Eq. \eqref{dustacceqSmin}.  

\begin{figure}
\centering
\resizebox{\hsize}{!}{\includegraphics[bb=255 10 510 260]{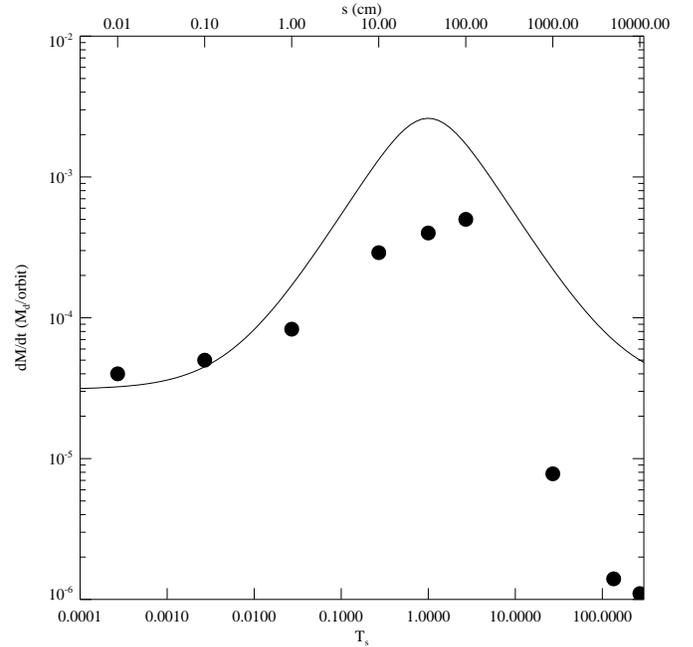}}
\caption{Accretion of particles onto a 0.1 \mjups planet.The solid line
gives the accretion rate of dust particles predicted by Eq. 
\eqref{eqMdotrel}.}
\label{dustaccfig11}
\end{figure}

From Fig. \ref{dustaccfig7} it is clear that essentially \emph{all} particles 
that make it through the resonances are eventually accreted by the planet. 
This is not true in general \citep{1993Icar..106..288K}, and especially
for low-mass planets a significant fraction of particles that cross the 
orbit of the planet are transferred to the inner disk rather than to be 
accreted onto the planet. This is illustrated for the case of $q=10^{-4}$ in 
Fig. \ref{dustaccfig11}. Moving from the smallest towards the largest particles
(from left to right in Fig. \ref{dustaccfig11}) we see three classes of 
particles:
\begin{itemize}
\item{I. Particles with $s<1$ cm are very strongly coupled to the gas, which
makes their radial velocity low enough for them to always reach the planet.}
\item{II. Particles with $s>1000$ cm are again trapped in exterior resonances,
and the accretion rate is therefore very low.}
\item{III. In between, there is a steady flow of particles onto the planet but
there is a significant fraction of particles that misses the planet.}
\end{itemize}
The transition from Class III to Class II happens again approximately at the
size predicted by Eq. \eqref{dustacceqSmin} (see also Fig. \ref{dustaccfig10}).
The deviation from the accretion rate given by Eq. \eqref{eqMdotrel} for 
Class III particles seems to increase with $s$, except for the largest Class 
III particle at $s=100$, which has a relatively high accretion rate. 

From Figs. \ref{dustaccfig7} and \ref{dustaccfig11} we conclude that particles 
that are trapped in exterior resonances are excluded from accretion onto the 
planet. Only particles smaller than $s_\mathrm{min}$ may reach the planet, 
where $s_\mathrm{min}$ is approximately given by Eq. \eqref{dustacceqSmin}. 
Particles with for which $1$ cm $ \le s \le s_\mathrm{min}$ potentially have a 
higher accretion rate than the gas. 

\subsection{Evolving gas disk}
However, when planets get massive enough to attract significant amounts 
of gas and dust they also start to reshape the surrounding disk. Planets as
massive as Jupiter open up a deep gas gap in the disk, but already 
Neptune-class planets restructure the gas disk in such a way that a \emph{dust}
gap emerges, even though there is no gap in the gas 
\citep{2004A&A...425L...9P}. It was also shown that this leads to a dramatic
decline in dust accretion \citep{paard07} for particles larger than $0.1$ cm. 
In this section we want to study the general case of dust accretion onto 
high-mass planets.

\begin{figure*}
\includegraphics[width=\textwidth]{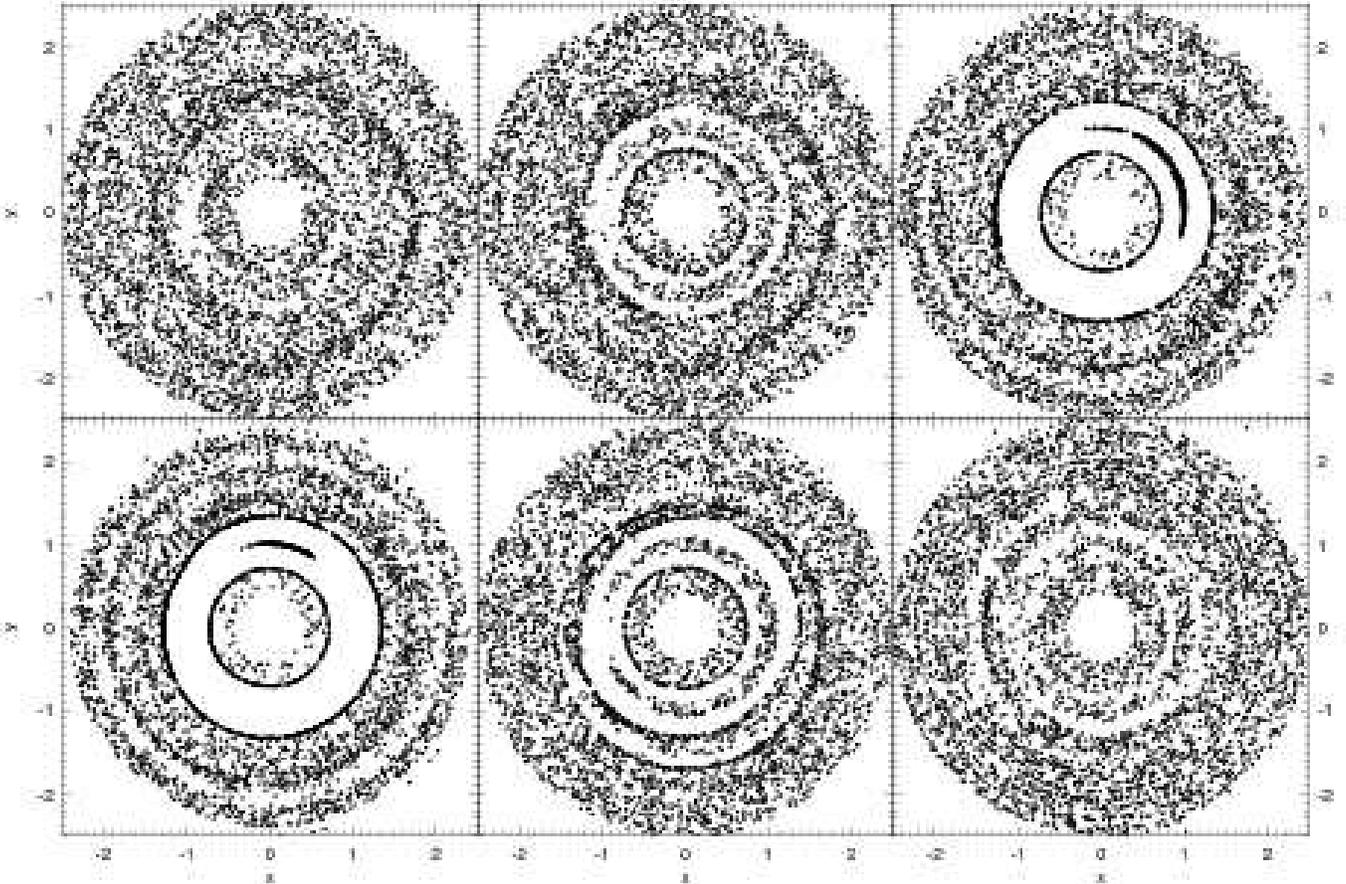}
\caption{Particle distribution after 20 orbits of a 1 \mjups planet for 
6 different particle sizes. Top row, from left to right: $s=0$ (perfect
coupling, $T_\mathrm{s}=0$), $s=1.0$ cm ($T_\mathrm{s}=0.027$) and $s=10.0$ cm 
($T_\mathrm{s}=0.27$). Bottom row, from left to right: $s=37.0$ cm 
($T_\mathrm{s}=1.0$), $s=1000.0$ cm ($T_\mathrm{s}=27.0$) and $s=\infty$ 
(no coupling, $T_\mathrm{s}=\infty$).}
\label{dustaccfig12}
\end{figure*}

We start by inspecting the particle distribution after 20 orbits of a 1
\mjups planet in Fig. \ref{dustaccfig12}. From the upper left panel to the 
lower right panel the stopping time increases from $T_\mathrm{s}=0$ (perfectly 
coupled particles, see Sect. \ref{dustaccsecPerfect}) to $T_\mathrm{s}=\infty$ 
(uncoupled particles, see Sect. \ref{dustaccsecNocoup}). After 20 orbits the 
planet did not get a chance
to open up a gap in the gas yet, as is clear from the upper left panel. 
The uncoupled particles in the lower right panel also do not clear a whole
gap, but instead two empty ''rails'' (see also Fig. \ref{dustaccfig4}).

For particles with $T_\mathrm{s}=0.027$ (upper-middle panel in Fig. 
\ref{dustaccfig12}) we see that the pressure gradients at the edges of the 
forming gap are pushing the particles away from the planet's orbit, as was 
also observed in \cite{2004A&A...425L...9P} for a ten times less massive 
planet. Because of the high mass of the planet in Fig. \ref{dustaccfig12}, 
and the corresponding large pressure gradients, the effect takes place
on much shorter time scales. Note also that because of the forming gap the
gas density is already significantly lower near the orbit of the planet, and
consequently the particles couple not as well to the gas as at the start
of the simulation. This, together with the larger pressure gradients at the
gap edges accounts for the fast evolution of the dust component of 1 cm.

When the stopping time approaches 1 the drift velocity of the particles
becomes so large that already after 20 orbits of the planet an almost empty
gap has formed (upper right and lower left panel of Fig. \ref{dustaccfig12}). 
The particles accumulate at the gap edges and at corotation. The latter because
when a gas gap is forming the density at $r=1$ is always larger than at
$|r-1|=0.1$ \citep[see][]{1999MNRAS.303..696K,paard06}, and the corresponding
pressure gradient pushes the dust particles towards $r=1$. However, when
the gas gap is sufficiently clean (which happens after approximately 100
orbits) this pressure gradient disappears and the particles near $r=1$ will 
be dragged inward.

When $T_\mathrm{s} > 1$ (lower middle panel in Fig. \ref{dustaccfig12}) we see 
again that the pressure gradient associated with the edges of the gas gap 
stimulate the formation of the two empty ''rails'' compared to the case of no 
coupling. The first signs of resonant capture can be seen as a dark eccentric 
ring outside the planet's orbit. Because these particles do not couple well to 
the gas it becomes progressively harder to clean the corotation region while
the gas gap is forming. Particles larger than $1000$ cm are not removed 
from the corotation region before the gas gap is complete, which makes the
horseshoe-shaped feature at $r=1$ almost permanent just as in the case of no 
coupling.

\begin{figure}
\centering
\resizebox{\hsize}{!}{\includegraphics[bb=260 10 495 240]{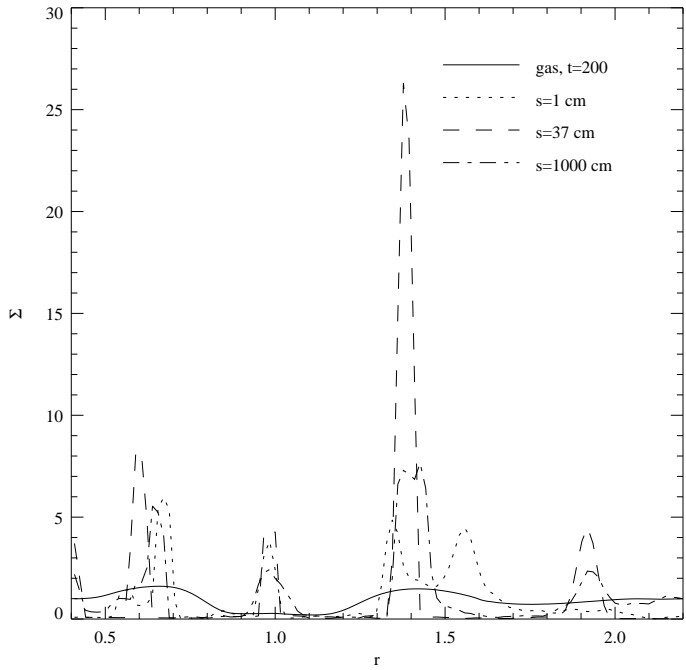}}
\caption{Azimuthally averaged surface density of gas and dust after $100$ 
orbits of a 1 \mjups planet for three different particle sizes.}
\label{dustaccfig13}
\end{figure}

At later times, resonant trapping comes to dominate the density structure
of the largest particles. In Fig. \ref{dustaccfig13} we show the azimuthally
averaged surface density after 100 orbits of the planet for the gas and
for three different particle sizes. The first thing that catches the eye is the
large peak in the density of particles with $s=37$ cm. The gas density
(solid line in Fig. \ref{dustaccfig13}) reveals that the pressure gradients 
around the peak all point \emph{away} from $r=1.4$. This makes this location an
efficient dust trap for basically all material originally located at
$1.1 < r < 1.8$. Locally the surface density is increased with a factor of
25.

The smaller particles (1 cm) move slower with 
respect to the gas than the $37$ cm particles. As a result two peaks can be 
seen near $r=1.4$: one from outward moving particles coming from $r\approx 1$ 
and one from inward moving particles coming from $r>1.7$. The large particles 
of 1000 cm also show a double-peaked feature. Note that the density gradient
near the 2:1 MMR at $r=1.6$ is such that the particles are dragged inward
more efficiently than in the case of the steady gas disk. Therefore while
Fig. \ref{dustaccfig10} predicts that the 1000 cm particles should be captured 
in the 2:1 MMR they are able to move all the way to $r\approx 1.4$.

\begin{figure}
\centering
\resizebox{\hsize}{!}{\includegraphics[bb=260 10 495 240]{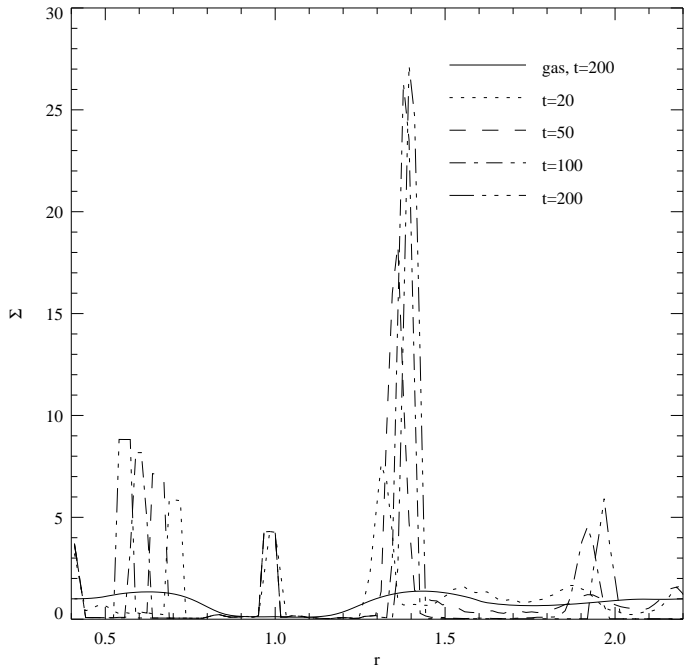}}
\caption{Particle surface density ($s=37$ cm) around a 1 \mjups planet
at different times.}
\label{dustaccfig14}
\end{figure}

In Fig. \ref{dustaccfig14} we show the time evolution of the surface
density of 
particles of $37$ cm. Four distinct features can be distinguished at
$r\approx 0.6$, $r\approx 1$, $r\approx 1.4$ and $r\approx 1.9$, which
are all due to the density structure in the disk. We have already mentioned
the large peak near $r=1.4$, which builds up largely from the 
inside within 100 orbits. Note that the peak slowly moves outward but 
stops near $r=1.4$, because it has arrived at a pressure maximum. 

\begin{figure}
\centering
\resizebox{\hsize}{!}{\includegraphics[bb=260 10 495 240]{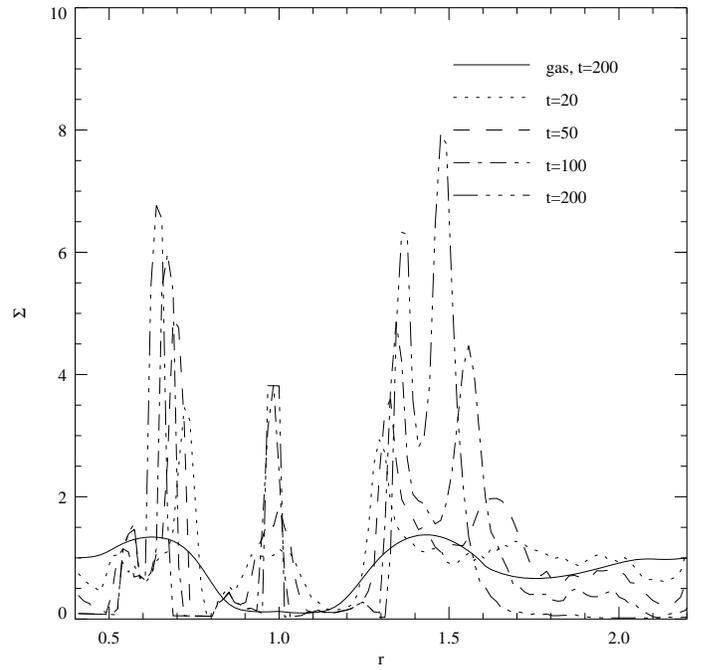}}
\caption{Particle surface density ($s=1$ cm) around a 1 \mjups planet
at different times.}
\label{dustaccfig15}
\end{figure}

For the peak near the inner edge of the gas gap at $r=0.6$ the situation is
slightly different. Again, the peak is moving slowly but this time the 
density gradient inside $r=0.6$ is not strong enough to stop the particles
completely. The peak slows down, but continues steadily towards the inner 
boundary and in the end it will move off the grid.

The peak at corotation is steady. Because the gas density has dropped
approximately 2 orders of magnitude, the effective stopping time at $r=1$
has increased from $1.0$ to $100.0$. This means that it will take several
hundreds of orbits more to remove this feature. Finally, a small outward
moving feature can be seen associated with the density gradient near $r=1.9$.
Note that resonant trapping is not important for particles of $s=37$ cm,
which agrees with Fig. \ref{dustaccfig7}. However, the density
gradients induced by the planet are so large that even particles of
$1000$ cm are not captured in the 2:1 MMR. 

The behavior of particles of $1$ cm is shown in Fig. \ref{dustaccfig15}.
First thing to note is the absence of the density feature near $r=1.9$, which
is due to the stronger gas drag that makes the particles follow the inward 
viscous motion of the gas. Near $r=1.4$ we can clearly see the two density
peaks originating from both sides of the gas density maximum slowly moving
towards each other. In contrast with the $37$ cm particles from Fig. 
\ref{dustaccfig14} the outer peak is the strongest, which is due to the lacking
feature near $r=1.9$. Particles of $37$ cm originally located at $r=1.8$ 
move outward, while their $1$ cm counterparts move inward and contribute to 
the inward moving peak near $r=1.4$.

\begin{figure}
\centering
\resizebox{\hsize}{!}{\includegraphics[bb=242 10 500 241]{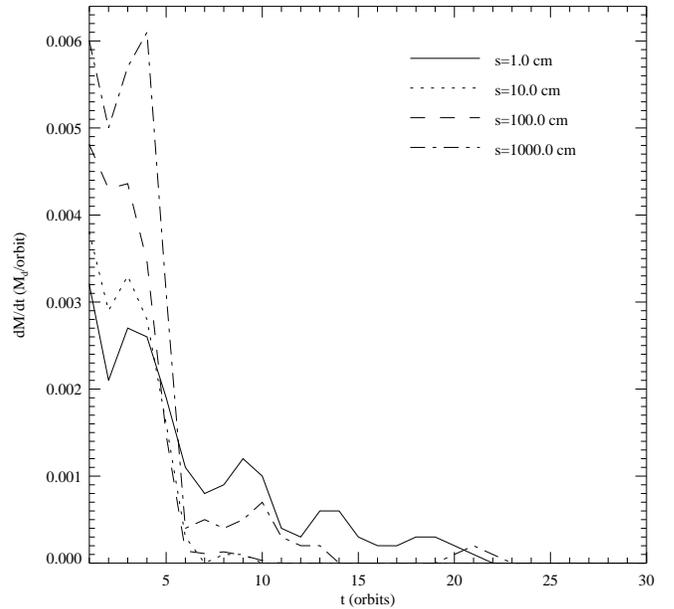}}
\caption{Accretion of particles onto a 1 \mjups planet and an evolving gas 
disk for four different particle sizes. After 30 orbits, all accretion rates 
stay exactly zero.}
\label{dustaccfig16}
\end{figure}

From Fig. \ref{dustaccfig12} we conclude that the particles that had
the largest accretion rate in Fig. \ref{dustaccfig7} actually create
the cleanest dust gaps. 
Already after 20 orbits the feeding zone is almost empty, and this has
severe implications for the accretion rates. In fact, for all particles that
open dust gaps the accretion rate equals zero after a finite amount of time.

This is illustrated in Fig. \ref{dustaccfig16}, where we show the accretion rate
for four different particle sizes. After only 30 orbits, all accretion rates
stay exactly zero. Note that this is unlike all previous models: for perfectly
coupled particles as well as uncoupled particles there was still residual
accretion after 200 orbits (see Fig. \ref{dustaccfig5}). This shows how powerful 
pressure structure in the disk is when it comes to moving dust particles.

The time scale at which accretion stops depends on the stopping time. The 
shortest time scales correspond to particles of 10 and 100 cm, which have 
an initial stopping time of $T_\mathrm{s} \approx 1$. Smaller particles as well as
larger particles have a smaller drift velocity (see Eq. \eqref{dustacceqDrift}) and
therefore it takes longer for them to clear the dust gap. Eventually, for the
smallest particle sizes we expect to recover the gas accretion rate, while
for the largest particle sizes we expect some residual accretion. 

\begin{figure}
\centering
\resizebox{\hsize}{!}{\includegraphics[bb=255 10 510 260]{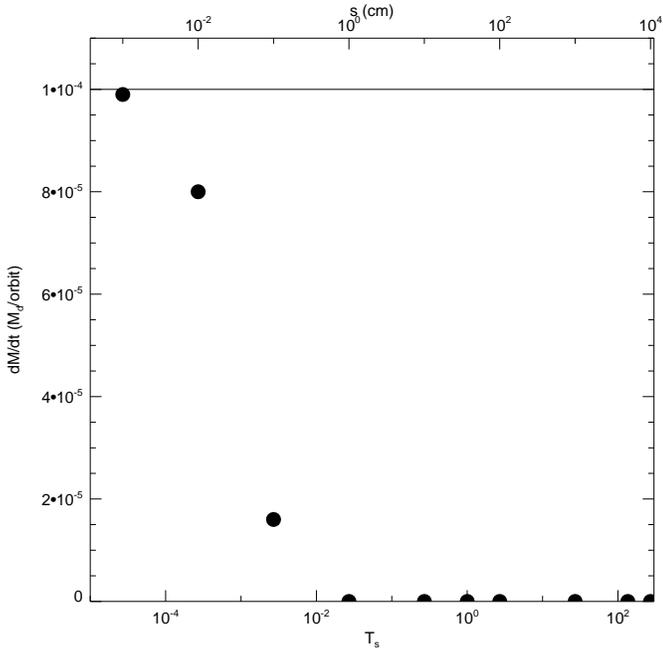}}
\caption{Accretion of particles onto a 1 \mjups planet and an evolving gas 
disk after 200 orbits. The solid line gives the gas accretion rate. All runs included 5000 
particles.}
\label{dustaccfig17}
\end{figure}

The accretion rate during the first 5 orbits of the planet grows 
monotonically with particle size. Particles of 1 cm accrete approximately as 
fast as perfectly coupled particles in this stage (see Fig. \ref{dustaccfig3}). 
Our resolution close to the planet is not high enough to see the Roche lobe 
clearing stage as was observed in \cite{paard07} using an adaptive mesh.
Larger particles accrete faster, and the particles of 1000 cm have an even 
higher accretion rate than uncoupled particles (see Fig. \ref{dustaccfig6}). The
turnover occurs around $T_\mathrm{s}=1$, which marks the transition from gas drag
dominated dynamics ($T_\mathrm{s}<1$) and gravity dominated dynamics ($T_\mathrm{s}>1$). During
this stage, when the planet has not opened up a gas gap yet, gas drag is able
to speed up accretion as long as we are in the gravity dominated regime. 
However, when the initial stopping time $T_\mathrm{s} < 1$ gas drag really takes over
and forces the particles to accrete at the same rate as the gas. 

In Fig. \ref{dustaccfig17} we show the accretion rate after 200 orbits of a 1 \mjups
planet for various particle sizes. Again we see that the accretion rate
is identically zero for a large range of sizes, while at both ends of the
figure we approach the accretion rates of the limiting cases of perfectly
coupled particles (left) and uncoupled particles (right). The
dust accretion rates almost never become steady, unless they are identically
zero or in the limiting case of perfectly coupled particles (cf. Figs. 
\ref{dustaccfig3} and \ref{dustaccfig5}). The radial distribution of the smallest particles
evolves on a time scale proportional to $T_\mathrm{s}^{-1}$, which becomes prohibitively
long when $T_\mathrm{s} \ll 1$. Therefore we chose a conservative approach to take
the accretion rate after 200 orbits, when the gas accretion has settled to
a steady state, as the final accretion rate for a given particle size, even
though the particle accretion rate is still declining. 

\begin{figure}
\centering
\resizebox{\hsize}{!}{\includegraphics[bb=255 10 510 260]{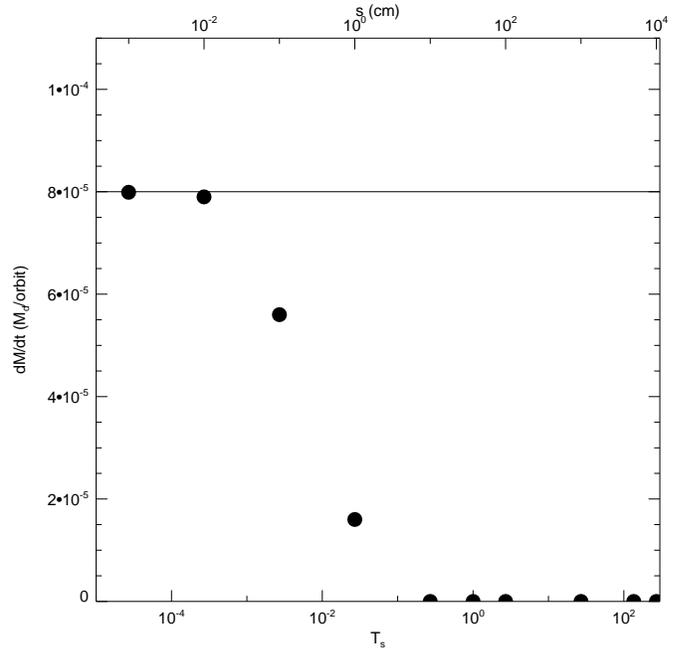}}
\caption{Accretion of particles onto a $0.1$ \mjups planet and an evolving gas 
disk after 200 orbits. The solid line gives the gas accretion rate. All runs included 5000 
particles.}
\label{dustaccfig18}
\end{figure}

The transition from no accretion to gas accretion begins around $s=0.001$ cm,
or equivalently $T_\mathrm{s}=0.000027$. This means that the total accretion flow onto
the planet is severely depleted in particles larger than $10$ \mum. Even more,
dust accretion never exceeds gas accretion, which makes it very difficult to
enrich a giant planet in solids.

A ten times less massive planet of $0.1$ \mjups does not open up a gas gap,
and therefore the planet-induced pressure gradients in the disk are less
strong. However, \cite{paard07} showed that even for such a small planet
dust accretion severely slows down. Figure \ref{dustaccfig18} shows the dust accretion
rates for various particle sizes. Again, we see accretion rates that are 
identically zero for $s>1.0$ cm, but the transition from gas accretion rates
to zero accretion is shifted to larger $s$ with respect to the planet of 1 
\mjup. We find that particles larger than 100 \mums have significantly low dust
accretion rates compared to the gas, which agrees with the result of
\cite{paard07}. Also for this low-mass planet it is very hard to get enriched
in solids.

More massive planets than Jupiter open up wider and deeper gaps, but the gap 
edges do not get much steeper. Therefore the minimum particle size that 
is able to move across the outer gap edge to be accreted by the planet is
the same as for the 1 \mjups planet. The width gas gap as well as the dust gap 
increases, though, which brings the density peak at the outer gap edge close
to the 2:1 MMR. This could lead to interaction between large bodies trapped
in the 2:1 MMR and smaller bodies trapped at the gap edge.

  
\section{Discussion}
\label{dustaccsecDisc}
When \cite{1985Icar...62...16W} discovered that orbital resonances are 
effective dust traps, they argued that this would not stop dust accretion
completely. The increase in eccentricity near the resonance would promote
collisions between planetesimals, and the resulting debris would be small
enough to move through the resonances towards the planet. 
\cite{1993Icar..106..288K} showed that this is not the whole story: 
particles small enough to make it through the orbital resonances have a 
significant chance of being transferred to interior orbits rather than 
being accreted onto the planet. We have demonstrated that there is yet
another barrier for smaller dust particles when the planet is massive enough
to induce significant pressure gradients in the disk, which is already 
the case for a planet of 15 \mearths \citep{paard07}.

According to Fig. \ref{dustaccfig10}, when a planetesimal breaks up into 100 cm
sized boulders in a catastrophic collision, these can subsequently be accreted
by a 1 \mjups planet. However, inspection of Fig. \ref{dustaccfig17} reveals that
100 cm boulders are stopped at the outer gap edge, and they need to be grinded 
down to 10-100 \mums before they can reach the planet. Moreover, it becomes
progressively harder to break smaller particles because they are better 
coupled to the gas and their relative velocities are low. This means
that the accretion flow onto the planet will be severely depleted in solids.
The same holds for a $0.1$ \mjups planet, although somewhat larger particles
may enter the accretion flow (see Fig. \ref{dustaccfig18}).

This has severe consequences for the final composition of the planet. According
to \cite{1996Icar..124...62P}, a 1 \mjups planet at $5.2$ AU approximately
spends the final $500,000$ years of its formation accreting the major part of
its massive gaseous envelope. Before that happened, the planet had already
reached a mass of $0.1$ \mjups, which makes it vulnerable to dust gap 
formation. Therefore the major part of the mass accreted by the planet is
relatively dust-poor compared to the interstellar value, which is of 
importance for the final composition of the planet. When Jupiter starts 
accreting gas when its solid core as well as the gaseous envelope both have a 
mass of 10 \mearths the difference in solid content between dust-rich gas
accretion and dust-poor gas accretion can be as large as 33 \%. Therefore, the enrichment in solids in Jupiter requires traditional enrichment scenarios (e.g. core erosion \citep{2004jpsm.book...35G}, clathrate hydrate trapping \citep{2001ApJ...550L.227G,2001ApJ...559L.183G}) should be much more effective than assumed until now.

\cite{1987Icar...70..319P}, in contrast with the suggestion by 
\cite{1985Icar...62...16W}, suggested that the accumulation of bodies in 
orbital resonances may lead to enhanced growth of planetesimals. Especially
for a planet on an eccentric orbit, for which the eccentricities of the 
planetesimals in low-order resonances stay rather low. In either case, the
size distribution of trapped particles may be very dynamical, because growth
as well as destruction is enhanced with respect to the rest of the nebula.
This makes it hard to predict the outcome of resonance trapping with respect
to the resulting size distribution.

Another interesting question is what the change in opacity does to the 
accretion flow. Dust is the major source of continuum radiation in the disk, 
and if the dust content is very low the gas may not be able to cool efficiently
through radiation. This may severely slow down gas accretion. The magnitude
of the opacity drop depends on how many of the smallest dust particles enter
the accretion flow, which in turn depends on the details of the collisions
of the larger bodies.

We have focused on planets of relatively high mass, in order for the two-
dimensional approach to be valid. For planets more massive than $0.1$ \mjups
the gas accretion rates in two and three dimensional simulations agree
reasonably well \citep{2003ApJ...586..540D}, and because the dust particles
are probably confined to an even thinner layer we believe that 
three-dimensional effects will be of minor importance. Also the results 
presented here do not depend on the density structure within the Roche
lobe of the planet.

Turbulence in circumstellar disks is probably of magnetic origin 
\citep{1990BAAS...22.1209B}. If strong enough, it may prevent the formation
of a deep gap but for reasonable disk parameters the gap formation paradigm
is not changed \citep{2004MNRAS.350..829P}. Note that the only thing that
is needed to trigger dust gap formation is the existence of planet-induced
pressure gradients, and therefore as long as the turbulent disk structure
does not overcome gap formation the mechanism presented here will operate
to reduce dust accretion onto high-mass planets.

The effect of a nonzero eccentricity of the planet's orbit on resonance 
trapping was studied by \cite{1987Icar...70..319P} and 
\cite{1995Icar..117....1K}. The effect on the gas disk, in particular the
formation of a gap, is not different from case of a circular orbit
\citep{2001A&A...366..263P}. Moreover, eccentricity growth of embedded 
planets is associated with deep annular gaps \citep{1992PASP..104..769A,
1993prpl.conf..749L,2003ApJ...585.1024G,2004ApJ...606L..77S,
2006A&A...447..369K}. Therefore it is likely that dust accretion onto 
eccentric planets also suffers from the planet-induced pressure gradients
in the disk.

In this paper, we have not considered migrating planets. Gap-opening planets experience Type II migration, for which the migration speed is comparable to the viscous inward drift of the gas. This migration rate is not enough to change the results described in the previous section. However, for planets that open up a shallow gas gap, the inward migration speed can be much higher. It is then possible that, depending on their size, particles will approach the planet both from the inner disk and the outer disk. The planet will overtake the smallest particles, while the particles that are less well-coupled to the gas still overtake the planet on their way in. This means that dust trapping in the inner resonances becomes possible for particles that are being overtaken by the planet.  Furthermore, the inner gap edge acts as a dust barrier for particles coming from the inner disk in the same way as the outer gap edge does for particles coming from the outer disk. Therefore, also in the case of a migrating planet dust accretion can be slowed down significantly.

When a planet is slowly accreting gas and dust from the surrounding disk, it will very gradually approach the mass that is required to open up a gap, first only in the dust disk, later also in the gas disk. Before that time, only particles that are captured in resonances will not reach the planet. This applies to the largest particles. As the planet gains mass, it may capture smaller and smaller particles in resonances. Therefore, the maximum particle size that may accrete onto the planet is decreasing with planetary mass (see also Figs. \ref{dustaccfig17} and \ref{dustaccfig18}).  

For simplicity, we have chosen a constant initial surface density. When the density profile is steeper, the surface density in the outer disk will be lower than in the case considered here. Therefore, particles in the outer disk will be less well coupled, the effects described in the previous section will be even more pronounced.

  
\section{Summary and conclusion}
\label{dustaccsecCon}
In this paper we have studied the accretion rates of solid particles of 
different sizes onto high-mass planets using a new particle-based method
for the integration of the equations of motion of the dust component. We have
shown that we can reproduce the gas accretion rates as found in hydrodynamical
simulations in the limit of perfectly coupled particles. In the limit of no
coupling to the gas, the planet is able to clear its feeding zone in 
approximately 100 orbits, regardless of the initial eccentricity distribution
of the particles. In order for accretion to proceed it is necessary to 
continuously replenish the feeding zone.

For a non-evolving gas disk three different classes of particles can be
distinguished: the smallest particles couple very well to the gas and they
accrete at the same rate as the gas, the largest particles get trapped in 
exterior resonances with the planet, and only particles with intermediate sizes
($1 $ cm $ \le s \le s_\mathrm{min}$) accrete onto the planet at a higher rate
than the gas.

When the gas is allowed to evolve dynamically it turns out that the 
planet-induced pressure gradients due to the formation of an annular surface
density depression play a major role in the movement of dust particles
towards the planet. Only the particles that couple very well to the gas 
are able to move across the outer edge of the density depression, which
limits the size of particles that can be accreted by a 1 \mjups planet
to $s \le 10$ \mum. For planets of lower mass somewhat larger particles
may be accreted, but even for a $0.1$ \mjups planet $s \le 100$ \mum.
The lack of accreting large bodies may have important consequences for 
growing planets in disks, especially with regard to their enrichment
in solids. If a large mass fraction of the solid component of the disk
resides in particles larger than $10$ \mums then it is not possible to 
enrich the planet.

\bibliographystyle{aa} 
\bibliography{6326.bib}

\end{document}